\documentclass[12pt]{iopart}
\usepackage{graphicx}
\usepackage{hyperref}
\usepackage{epigraph}

\begin{document}
\eqnobysec
\title{On the foundations of quantum physics}
\author{E. A. Solov'ev}

\address{Bogoliubov Laboratory of Theoretical Physics, Joint Institute for Nuclear Research, 141980 Dubna, Moscow region, Russia}

\ead{esolovev@theor.jinr.ru}

\begin{abstract}
Some aspects of the interpretation of quantum theory are
discussed. It is emphasized that quantum theory is formulated in a
Cartesian coordinate system; in other coordinates the result
obtained with the help of the Hamiltonian formalism and commutator
relations between 'canonically conjugated' coordinate and momentum
operators leads to a wrong version of quantum mechanics. In this
connection the Feynman integral formalism is also discussed. In
this formalism the measure is not well-defined and there is no
idea how to distinguish between the true version of quantum
mechanics and an incorrect one; it is rather a mnemonic rule to
generate perturbation series from an undefined zero order term.
The origin of time is analyzed in detail by the example of atomic
collisions. It is shown that the time-dependent Schr\"odinger
equation for the closed three-body (two nuclei + electron) system
has no physical meaning since in the high impact energy limit it
transforms into an equation with {\it two independent} time-like
variables; the time appears in the {\it stationary} Schr\"odinger
equation as a result of extraction of a classical subsystem (two
nuclei) from a closed three-body system. Following the
Einstein-Rosen-Podolsky experiment and Bell's inequality the wave
function is interpreted as an actual field of information in the
elementary form. The relation between physics and mathematics is
also discussed.
\end{abstract}

\epigraph{Eine neue wissenschaftliche Wahrheit pflegt sich nicht
in der Weise durchzusetzen, dass ihre Gegner {\"u}berzeugt werden
und sich als belehrt erkl{\"a}ren, sondern dadurch, dass die
Gegner allm{\"a}hlich aussterben und dass die heranwachsende
Generation von vornherein mit der Wahrheit vertraut gemacht
ist.(In English: A new scientific truth does not triumph by
convincing its opponents and making them see the light, but rather
because its opponents eventually die, and a new generation grows
up that is familiar with it.)}{M.Planck}

\section{Introduction}\label{intro}

This paper is an extended version of my talk given at the
International Conference "Symmetries in Physics" dedicated to the
90th anniversary of Ya.A. Smorodinsky (March 27 - 29 2008, Dubna,
Russia) \cite{Sol09}, where some general aspects in the
interpretation of quantum theory are reviewed. This topic has an
eighty-year history and many versions have been proposed,
beginning with the Copenhagen interpretation. Here three new
points are presented:

$\bullet$ Quantum theory is much less symmetric than classical
mechanics. In contrast to classical mechanics, quantum theory is
not covariant with respect to canonical transformations. Moreover,
an analogue of  a canonical transformation cannot be introduced into
quantum mechanics.

$\bullet$ The time-dependent Schr\"odinger equation for closed
systems is an artifact and has no fundamental meaning; the time
appears in the stationary Schr\"odinger equation as a result of
extraction of a classical subsystem from a total system.

$\bullet$ Quantum theory, in its origin, deals with a new object -
"field of information" which is additional to the objects
"material point" introduced in classical mechanics and
"electromagnetic field" introduced in classical electrodynamics.

All these statements are not a personal opinion - they are proved
at the standard physical level.

Within the last point the role of the concept  of 'measurement' and
its connection with the way we take a decision (or how our brain
acts) are analyzed. In the conclusion, the relation between physics
and mathematics is also discussed.

Here the philosophical and popular science papers related to this
topic are not discussed since they are not based on the laws of
nature but on the metaphysical assumptions which are beyond the
natural-science standards. A very nice description of such
approach one may find in  Feynman's book \cite{Fey0}, Chapter
'Cargo Cult Science'.

\section{Symmetry in quantum and classical physics}
\label{sec:1}

The Schr\"{o}dinger equation is given in the Cartesian coordinate
system $\{x,y,z\}$
\begin{equation}
\Bigl[-\frac{\hbar^2}{2m}\Bigl(\frac{\partial^2}{\partial
x^2}+\frac{\partial^2}{\partial y^2}+\frac{\partial^2}{\partial
z^2}\Bigr) +V(\bi{r})\Bigr]\Psi_n(\bi{r})=E_n\Psi_n(\bi{r}).
\label{2.-5}
\end{equation}
The exclusive role of Cartesian coordinates is quite well-known
(see, e.g., \cite{Fock31}). Equation (\ref{2.-5}) is a law of
nature and it cannot be derived. In all other coordinates (or
representations) the result is obtained by a mathematical
transformation of equation (\ref{2.-5}) from the Cartesian
coordinates to desirable coordinates (or representation) which, of
course, does not change the physical (observable) results. An
alternative way to introduce the Schr\"odinger equation in
arbitrary coordinates based on 'canonical' transformations has
serious problems. Generally, the momentum operator $\hat p$,
defined in the $q$-representation by the commutator relation
$q\hat p - \hat p q=i\hbar$, has the form
\begin{equation}
\hat p=-i\hbar \frac{\partial}{\partial q}+f(q).
\end{equation}
The function $f(q)$ is determined by Hermiticity condition in a
given coordinate system
\begin{equation}
\int \psi_1^*(q) \hat p \psi_1(q) J(q)dq=\int [\hat p \psi_1(q)]^*
\psi_1(q) J(q)dq
\end{equation}
from which
\begin{equation}
\hat p=-i\hbar \frac{1}{\sqrt{J(q)}}\frac{\partial}{\partial
q}\sqrt{J(q)},
\end{equation}
where $J(q)$ is  the Jacobian. In the spherical coordinates
$\{r,\vartheta,\varphi\}$ the 'canonically conjugated' Hermitian
components of the momentum operators are
\begin{equation}
\hat{p}_{r}=-i\hbar\frac{1}{r}\frac{\partial}{\partial r}r , \ \ \hat{p}%
_{\vartheta}=-i\hbar\frac{1}{\sqrt{\sin \vartheta}}
\frac{\partial}{\partial
\vartheta}\sqrt{\sin \vartheta}, \ \ \hat{p}_{\varphi}=-i\hbar \frac{\partial}{%
\partial \varphi} .  \label{2.-2}
\end{equation}
However, if we substitute $\hat{p}_{r}, \hat{p}_{\vartheta}$ and $\hat{p}%
_{\varphi}$ in the operator of kinetic energy, an incorrect result
is obtained:
\begin{equation}
\hat{T}=\frac{1}{2m}\Biggl[\hat{p_r}^2+\frac{1}{r^2}\Biggl( \hat{p_{\vartheta}}%
^2+\frac{\hat{p_{\varphi}}^2}{\sin^2\vartheta}\Biggr)\Biggr]= -\frac{\hbar^2}{%
2m}\Biggl(\Delta+\frac{\cos^2\vartheta-2}{%
4r^2\sin^2\vartheta}\Biggr),  \label{2.-a}
\end{equation}
since a term additional to the Laplacian appears, i.e., in the
simplest case of transformation from the Cartesian to spherical
coordinates the Hamiltonian formalism (which is mostly
coordinate-momentum variables plus covariance of the theory with
respect to the canonical transformation) does not work. Thus, the
momentum operator, as well as the Hamiltonian formalism, is not
well defined.

The incompatibility of the canonical transformations with quantum
theory becomes obvious if we consider the Bohr-Sommerfeld
quantization rule
\begin{equation}
\int_{q_1}^{q_2}p(q)dq=\pi\hbar(n+\alpha_1+\alpha_2),
\end{equation}
where $q_i$ is a turning point (caustic, which is a {\it
non-local} characteristic of trajectory) and $\alpha_i$ is a
quantum phase shift in it. Here the left-hand side is invariant
with respect to the canonical transformation whereas the
right-hand side changes since caustics can disappear under
canonical transformation, for instance, after transformation to
action-angle variables. In \cite{Rich}, the action-angle variables
are used to obtain the semiclassical energy spectrum for the
quadratic Zeeman effect. However, in these variables the
information about caustics is absent, and the quantization condition
(3.8) in \cite{Rich} erroneously employs an integer quantum number
instead of semi-integer which leads to a wrong energy spectrum.
The correct semiclassical energy spectrum is obtained in
configuration space \cite{Sol82}. Thus, the canonical
transformations are not compatible even with the Bohr-Sommerfeld
quantization condition.

Many textbooks ignore these facts treating the commutator relation
between coordinate and generalized momentum operators as a
background of quantum mechanics (see, e.g., \cite{Dir}) or
describing the states by the Feynman integral (see, e.g.,
\cite{Fey}). In particular, the Feynman integral cannot be
rigorously determined since the measure of this integral does not
exist as a mathematical quantity. From the physical point of view
the measure cannot be introduced since in this approach {\it there
is no idea} how to distinguish the true version of quantum
mechanics from an incorrect one; it is rather a mnemonic rule to
generate perturbation series from an undefined zero order term.
All these treatments provoke a wrong opinion that quantum theory
has the same level of symmetry as classical mechanics.

\section{Stationary and time-dependent
Schr\"odinger equation} \label{sec:2}

In 1926 Schr\"odinger \cite{Sch1} formulated quantum theory as an
eigenvalue problem for the Hamiltonian operator
\begin{equation}
\hat{H} \Psi_n= E_n\Psi_n. \label{1}
\end{equation}
The validity of this representation was confirmed, e.g., by
comparison of the obtained energy spectrum of the hydrogen atom
with experimental data. In \cite{Sch2}, Schr\"odinger introduced
the time-dependent equation
\begin{equation}
\hat{H} \Psi= i\hbar\frac{\partial}{\partial t}\Psi. \label{2}
\end{equation}
However, the Hamiltonian operator does not contain the time, and
equation (\ref{2}) looks like an artificial
construction\footnote{For instance, initially Schr\"odinger
introduced an equation of second order in the time derivative.}.
Besides, from the general point of view,  time is not observable
in quantum theory - there is no time-like Hermitian operator. Some
aspects of this problem were discussed, e.g., in \cite{Sol1} and
\cite{Briggs}. Here the emergence of time in atomic collision
theory is analyzed in detail.

The inelastic transitions of  an electron in an atomic collision are
described in the c.m. frame of reference by the Schr\"odinger
equation
\begin{equation}
\hat{{\cal H}} \Psi ({\bi R},{\bi r}) = {\cal E} \Psi ({\bi
R},{\bi r}) , \label{3}
\end{equation}
where
\begin{equation}
\hat{{\cal H}}=-{\hbar^2 \over {2M}} \Delta_{\bi R}+{Z_1Z_2\over
R}+\hat H_{el.} ({\bi R}) \nonumber
\end{equation}
is the three-body Hamiltonian operator, $Z_1, Z_2$ are the charges
and $M$ is the reduced mass of the nuclei, $\bi R$ is the
internuclear radius-vector, $\bi r$ is the radius-vector of
electron referred to c.m. of nuclei, ${\hat H}_{el.}({\bi R})$ is
the electronic part of the total Hamiltonian. In the adiabatic
representation the wave function $\Psi ({\bi R},{\bi r})$ has the
form \footnote{It is assumed that the impact velocity is directed
along the $z$-axis.}:
\begin{equation}
\Psi ({\bi R},{\bi r})= \sum _{n,L} {F_n^{L}(R)\over
R}Y_{L0}(\theta,\varphi)\phi _n ({\bi r}; {\bi R}) , \label{5}
\end{equation}
where $Y_{LM}(\theta,\varphi)$ are spherical functions (angular
part of the nuclear wave function), $\phi _n ({\bi r};{\bi R})$
are the adiabatic wave functions, and $E_n (R)$ are the adiabatic
energies -
\begin{equation}
\hat H_{el.} (\bi R) \phi _n ({\bi r};\bi R) = E_n (R) \phi_n
({\bi r};\bi R) \label{6}
\end{equation}
which depend on $\bi R$ as an external parameter in the
Hamiltonian $\hat H_{el.} (\bi R)$. After the substitution of
(\ref{5}) into (\ref{3}), the Schr\"odinger equation takes the
form of a system of equations for the radial nuclear wave
functions
\begin{equation} \fl
\Bigl[-{\hbar^2\over {2M}} {d^2\over dR^2}+{{{\cal L}^2\over
{2{M}R^2}}}+{Z_1Z_2\over R}+E_n(R)-{\cal E} \Bigl] F_n^{L}(R) =
\sum_{n'L'}  \hat W_{nn'}^{LL'}(R)F_{n'}^{L'}(R), \label{7}
\end{equation}
where ${\cal L}^2=\hbar^2L(L+1)$ is the squared nuclear angular
momentum. The operator of the non-adiabatic coupling $\hat
W_{nn'}^{LL'}$ is equal to
\begin{eqnarray}
&&\hat W_{nn'}^{LL'}=\frac{\hbar^2}{2M}\Bigl[\biggl(2 \biggl
\langle n\biggl|\frac{\partial}{\partial R}\biggr| n'
\biggr\rangle \frac{\partial}{\partial R}+\biggl\langle
n\biggl|\frac{\partial^2}{\partial
R^2}\biggr|n'\biggr\rangle+\frac{1}{R^2}\biggl\langle
n\biggl|\frac{\partial^2}{\partial
\theta^2}\biggr|n'\biggr\rangle\biggr)\delta_{LL'}+ \nonumber
\\
&&+\frac{2i}{\sqrt{2L+1}R^2}\biggl\langle
n\biggl|\frac{\partial}{\partial
\theta}\biggr|n'\biggr\rangle\biggl(\frac{L^2}{\sqrt{2L-1}}\delta_{L,L'+1}+\frac{(L+1)^2}{\sqrt{2L+3}}\delta_{L,L'-1}\biggr)\Bigr],
\label{7'}
\end{eqnarray}
where $\langle n|{\hat A}|n'\rangle=\int \phi_n(\bi r;\bi R) {\hat
A}\phi_{n'}(\bi r;\bi R)d\bi r$ is the matrix element of the
operator $\hat A$ between $|n\rangle$ and $|n'\rangle$ adiabatic
electronic states.

In the semiclassical limit ($\hbar \to 0$, $L\to \infty$) the
solution has an essential singularity which can be extracted
explicitly:
\begin{equation}
F_n^L(R) = \tilde{F}_n^L(R)e^{-\frac{i}{\hbar}S_n(R)} , \label{8}
\end{equation}
where $S_n(R)=\int\displaylimits^{R}{\cal P}_n(R') dR'$ is the
classical action and
\begin{equation}
{\cal P}_n(R)=\sqrt {2M\Bigl[{\cal E}-{{{\cal L}^2}\over
{2MR^2}}-{Z_1Z_2\over R}-E_n(R)\Bigl]} \label{9}
\end{equation}
is the radial momentum of the nuclei in the $n$-channel. In terms
of $\tilde{F}_n^L(R)$ the Schr\"odinger equation (\ref{3}) has the
form ($\Delta S_{nn'}(R)=S_n(R)-S_{n'}(R)$)
\begin{eqnarray}
i\hbar\frac{{\cal P}_n(R)}{M}{d\tilde{F}_n^{L}(R)\over dR}
-{\hbar^2\over {2M}} {d^2\tilde{F}_n^{L}(R)\over dR^2}=\sum_{n'L'}
e^{\frac{i}{\hbar}\Delta S_{nn'}(R)} \hat
W_{nn'}^{LL'}(R)\tilde{F}_{n'}^{L'}(R), \label{10}
\end{eqnarray}
Since $\tilde{F}_n^L(R)$ is a smooth function with respect to
$\hbar \to 0$, we can neglect the second term in the left-hand
side of this equation. In the right-hand side the dominant
contribution is obtained through differentiation of exponent
$e^{-iS_{n'}(R)/\hbar}$ over $R$ (first term in (\ref{7'})) and if
we neglect the difference between $L$ and $L\pm 1$ in the second
line of expression (\ref{7'}). In this limit the nuclear angular
momentum is conserved and equation (\ref{10}) takes the form
\begin{eqnarray} \fl
i\hbar\frac{{\cal P}_n(R)}{M}{d\tilde{F}_n^L(R)\over dR} =i\hbar
\sum_{n'} \biggl\langle n\biggl|\frac{{\cal
P}_{n'}(R)}{M}\frac{\partial}{\partial R}+\frac{{\cal L
}}{MR^2}\frac{\partial}{\partial \theta }\biggr|n'\biggr\rangle
e^{\frac{i}{\hbar}\Delta S_{nn'}(R)} \tilde{F}_{n'}^L(R).
\label{11'}
\end{eqnarray}

If ${\cal E}\gg E_n(R)$, the classical approximation for  nuclear
motion can be employed. Treating $E_n(R)$ as a small quantity, the
momentum ${\cal P}_n(R)$ can be approximated \cite{Sol1}: $${\cal
P}_n(R)\approx MV(R),$$ where $V(R)$ is the radial nuclear
velocity without $E_n(R)$. Then the factor $MdR/{\cal
P}_n(R)\approx dR/V(R)=dt$ has the meaning of the differential of
time, the exponent $\Delta S_{nn'}(R)\approx \int^t \Delta
E_{nn'}(R(t'))dt'$ ($\Delta E_{nn'}(t)=E_n(R(t))-E_{n'}(R(t))$)
and the Schr\"odinger equation takes the well-known time-dependent
form
\begin{eqnarray}
i\hbar{d\over dt} \tilde{F}_n^L(t)=i\hbar \sum_{n'} \biggl\langle
n\biggl|\frac{dR(t)}{dt}\frac{\partial}{\partial
R}+\frac{d\theta(t)}{dt}\frac{\partial}{\partial
\theta}\biggr|n'\biggr\rangle
\tilde{F}_{n'}^L(t)e^{-\frac{i}{\hbar}\int^t\Delta
E_{nn'}(t')dt'}=\nonumber
\\
=i\hbar \sum_{n'} \biggl\langle n\biggl|\frac{d}{d
t}\biggr|n'\biggr\rangle\tilde{F}_{n'}^Le^{-\frac{i}{\hbar}\int^t\Delta
E_{nn'}(t')dt'}, \label{11}
\end{eqnarray}
where the classical definition of angular momentum ${\cal
L}=MR^2d\theta(t)/dt$ was used. In \cite{Fock}, equation
(\ref{11}) was obtained from the time-dependent Schr\"odinger
equation
\begin{equation}
{\hat H}_{el.}({\bi R}(t)) \psi ({\bi r},t) =
i\hbar\frac{\partial}{\partial t} \psi ({\bi  r},t) \label{12a}
\end{equation}
after substitution wave function in the form
\begin{equation}
\psi ({\bi r},t)=\sum_{n}
\tilde{F}_{n}^L(t)e^{-\frac{i}{\hbar}\int^t E_{n}(t')dt'} \phi _n
({\bi r};{\bi R}(t)).
\end{equation}
Thus, the time-dependent Schr\"odinger equation (\ref{12a}) is
{\it derived} straightforwardly from the stationary Schr\"odinger
equation (\ref{3}). From the above considerations it is clear that
'time' can be introduced in the high impact energy limit only when
we can neglect the dependence of ${\cal P}_n(R)$ on $n$. At low
impact energy, the solution of the Schr\"odinger equation
(\ref{7}) is the superposition of nuclear motions in different
$n$-channels having significantly different momenta ${\cal
P}_n(R)$ and a unique time does not exist.

Now let us consider the time-dependent equation
\begin{equation}
\hat{{\cal H}} \Psi ({\bi R,\bi r,t}) =
i\hbar\frac{\partial}{\partial t} \Psi ({\bi R,\bi  r,t})
\label{12}
\end{equation}
instead of the stationary Schr\"odinger equation (\ref{3}). If we
apply the same approximation as above, we obtain an analogue of
equation (\ref{11}) which has {\it two independent time-like
variables}: one existing in the initial form of equation
(\ref{12}) and the other coming from the second derivative with
respect to $R$, i.e. this equation is meaningless. Just the
extraction of the classical subsystem from the total system in the
stationary Schr\"odinger equation leads to the replacement of the
coordinates of the classical subsystem by the time variable. It is
the sole reason for emergence of the actual time in the
Hamiltonian operator. This analysis plays the role of {\it
experimentum crucis} in the clarification of the place of the time in
quantum theory. In the simpler case of the two-body problem the
conflict between stationary (\ref{1}) and time-dependent (\ref{2})
equations is obscure since  inelastic channels do not exist and
we cannot extract a classical subsystem.

In the scattering theory (two-body problem) the time-dependent
Schr\"odinger equation is  employed to avoid the normalization
problem in the continuum (see, e.g., \cite{New}) - the stationary
problem is out of the Hilbert space and we have no well-defined
{\it physical} stationary states. It is a very serious defect of
the theory. Below the boundary of the continuum, the aim of the
theory is the search for stable (or stationary) states which exist
at discrete energy values only. Above the boundary of continuum
the problem is completely different. The concept 'stable' or
'unstable' does not exist here. The solution of the stationary
Schr\"odinger equation cannot be normalized to unit. In classical
approach the particle comes from infinity and then goes to
infinity as $t \to \infty$; probability to find it at any finite
distance is equal to zero and to find it at infinity is equal to
unit. Some tricky way to resolve this problem is to introduce the
wave packet formalism. However, as was shown above, the 'time'
variable in this formalism has no direct physical meaning; the
wave packet construction plays rather a psychological role. At the
same time, all physical results in the scattering theory are
derived from the stationary Schr\"odinger equation.

Generally, the presence of the time in the time-dependent
Schr\"odinger equation means that some subsystem is under our
non-stop control since the time is part of the frame of reference
which is a tool for measurement. The connection between two frames
of reference (Galileo or Lorenz transformation) is also a
classical concept which is beyond quantum theory. Quantum theory
describes a new object - the field of information (see the next
section) which has no relation to the time for closed systems.

\section{The information field and measurement}\label{sec:3}

In classical mechanics and classical electrodynamics the concepts
'material points' and 'electro-magnetic waves' were introduced. In
this sense we can consider quantum theory as a theory of an
'information field' (see, e.g.,\cite{Sol09}) which is a wave
function $\Psi(x,y,z)$. At first sight this statement looks
non-persuasive, but this case is similar to the situation with the
electro-magnetic field - nobody suspected its existence before
classical electro-dynamics. The interaction between 'material
points' and 'information field' is the act of measurement. Thus,
'information field' is an additional object to 'material points'
and 'electro-magnetic waves' but without mass and energy. This
statement is in contradiction with wave-particle duality. In the
present interpretation we have material points characterized by
the coordinates $x,y,z$, and the 'information field' $\Psi(x,y,z)$
which is subject to the wave-type Schr\"odinger equation. Thus,
there is no duality because of two different objects - the
material points and the information field. {\it Experimentum
crucis} proving the existence of the information field, is based
on the Einstein-Podolsky-Rosen phenomenon \cite{Ein} and Bell's
inequality \cite{Bell} which allows one to distinguish whether the
quantum approach is complete or not. One of the realizations of
this experiment is the following. After a radiative decay of a
hydrogen atom from a metastable $2S$-state two photons are emitted
having spins with opposite directions
\begin{equation}
{\bi s}_1=-{\bi s}_2, \label{3.1}
\end{equation}
since the initial ($2S$) and final ($1S$) states of hydrogen have
angular momentum equal to zero. However, the orientation of each
spin is uncertain in the same manner as the position of a particle
in a potential well. If the measurement of the direction of the
spin of the first photon is performed, the second photon takes the
opposite direction of spin, according to (\ref{3.1}),
independently of the distance between photons. The fixation of the
spin orientation of the second photon happens
immediately\footnote{This does not contradict the relativistic
restriction $v\leq c$ because the carrier of information is not
material, i.e., it has no relation to the transfer of mass or
energy.} and it is an actual changing of its state which is
confirmed by experiment \cite{Exp}, where it was found that the
speed of quantum information is at least $10^{4}$ greater than the
speed of light $c$. Measurement gives orientation of spin,
according to this field of information. Intuitively, it is clear
what the concept 'measurement' means but it cannot be formalized,
i.e., to be written in a mathematical form. The nontrivial role of
measurement was most clearly realized in quantum physics.

The fresh demonstration of the existence of a common information field
is the effect of long-range near-side angular correlations in
proton-proton collisions at the LHC in CERN \cite{CMS}.

The interpretation of the wave function as a common field of
information makes the well-known facts such as a collective
behavior of bees and ants, bird navigation, extremely fast baby's
mental development, d\'{e}j\`{a} vu phenomenon {\it etc.} more
transparent. Perhaps such a object (common field of information)
could clarify how our brain acts. Here the statement made by
Svyatoslav Medvedev (Director of the Institute of Human Brain,
Russian Academy of Sciences) in the TV program dedicated to the
memory of the Russian psychiatrist Vladimir Bekhterev can be
cited: " ... our brain is  the {\it interface} between the
material and the spiritual world ...". In this context, brain
activity can be interpreted as follows: consciousness prepares the
incoming information into an appropriate form to send it to
subconsciousness (common information field level). What happens in
subconsciousness is hidden from our mind, but the final decision
suddenly appears (sometimes after a long time - depending on the
level of the problem). This event is similar to the spin
measurement in the Einstein-Podolsky-Rosen experiment and can be
considered as the most fundamental form of 'measurement'. Here our
brain as an 'interface' plays the role of a tool of measurement.

Generally, we do not know so much about consciousness even with
respect to Socrates. Modern set of tools in this area such as
measurement of temperature, blood pressure, electroencephalography
{\it etc.} are taken from inorganic natural sciences. All of them
have no direct relation to conscious activity and memory; it is
the same as during the Hellenistic period to try to investigate
consciousness by means of a ruler and balance. That is why at the
end of her life Natalia Bekhtereva (Director of the Institute of
Human Brain 1990-2008) said\footnote{Interview in the Russian
newspaper "Rossiyskaya gazeta" from June, 26th 2003}: "Doesn't
exist yet not only theory, but even a reasonable hypothesis how
the brain works". In addition, a scientific approach is based on
the assumption that physical events are reproduced at any place
and any moment under the same external conditions. In this sense
consciousness as well as memory cannot be a subject of science
since it is personal and cannot be repeated. Besides, we have no
idea where they are located. Moreover, it seems that space and
matter are irrelevant here. The only vague way to contemplate the
consciousness is perhaps through analysis of our dreams which are
out of our autocontrol and instincts. That is why often we get the
decision during the dream.

\section{Physics and mathematics}\label{sec:4}
Physics as science was founded by Galileo Galilei. He discovered the
fundamental role of experiments in inorganic nature - physical
events are reproduced at any place and any moment under the same
external conditions, i.e. they obey the laws of nature which take
the form of mathematical equations introduced by Isaac Newton.
Thus, physical theories are based on mathematics which is a
logical scheme related rather to our mentality than to nature. It
is a miracle but mathematics works! Mathematics is based on a few
axioms such as $a+b=b+a$, $c(a+b)=ca+cb$, $(a+b)+c=a+(b+c)$ and so
on. In principle, mathematics does not produce new information;
its aim is the transformation, according to these axioms, of the
initial expression to a form more transparent for our
consciousness, e.g., by solution of a differential equation. Thus,
mathematics plays a subordinate, auxiliary role in physics.
Investigations based only on the abstract mathematical background
lead, sometimes, to artefacts.

The first example is the case of irregular motion in classical
mechanics. In spite of a huge number of papers devoted to this
problem there is no important achievement for real physical
systems. Irregular motion is mostly studied in 2D billiard
models and 1.5D models which are in fact non-Hamiltonian systems.
Interest in the irregular motion is explained by an attempt to
understand the origin of the arrow of time as provided by the Second Law
of Thermodynamics which says that in an isolated system entropy
tends to increase with time.  However, there is a fundamental
contradiction between such evolution and classical mechanics -
the first is 100\% irreversible in time whereas classical
mechanics is 100\% reversible. Thus, classical mechanics rather
deals with pseudo-chaotic evolution which is due to the
fundamental difference between rational and irrational numbers in
mathematics which does not exist in nature. Generally speaking, the
axioms of mathematics provide always reversible time behaviour.
We can obtain irreversible evolution only if we introduce a
stochastic concept such as 'probability' which is beyond standard
mathematics.

The next example is Gutzwiller's 'theory'. In this approach, the
contribution of the unstable periodic orbit to the trace of the
Green function is determined by the  formula \cite{Gutz}
\begin{equation}
g(E)\sim -\frac{iT(E)}{2\hbar}\sum_{n=1}^{\infty}\frac{\exp\{in[S(E)/\hbar-%
\lambda\pi/2]\}}{\sinh[nw(E)/2]}  \label{4.1}
\end{equation}
where $S(E)$, $w(E)$, $T(E)$ and $\lambda$ are the action, the
instability exponent, the period and the number of focal points
during one period, respectively. After the expansion of the
denominator, according to
$[\sinh(x)]^{-1}=2e^{-x}\sum_{k=0}^{\infty}e^{-2kx}$, and
summation of the geometric series over $n$ one can see that the
response function (\ref{4.1}) has poles at complex energies
$E_{ks}$ whenever
\begin{equation}
S(E_{ks})=\hbar\lambda\pi/2-i\hbar w(E_{ks})\left(k+\frac{1}{2}%
\right)+2s\pi\hbar,  \ \ k,s=0,1,2,...  \label{4.2}
\end{equation}
In fact, expression (\ref{4.1}) has been introduced {\it ad hoc};
it is based on local characteristics in the vicinity of the
periodic orbit ignoring the asymptotic region which is responsible
for the physical boundary condition. Thus, in this scheme it is
impossible to distinguish whether the energy belongs to the
discrete spectrum or the continuum. Besides, in nonseparable systems
unstable periodic orbits with long periods lie everywhere dense in
phase space (renormalization-group) and the response function
(\ref{4.1}) has a pathological structure like the Weierstrass
function which is continuous everywhere but differentiable
nowhere. The singularities predicted by expression (\ref{4.1})
have no physical meaning in the case of the discrete spectrum as well,
since the energies $E_{ks}$ are complex.

\begin{figure}
\begin{center}
\includegraphics[width=0.7\textwidth]{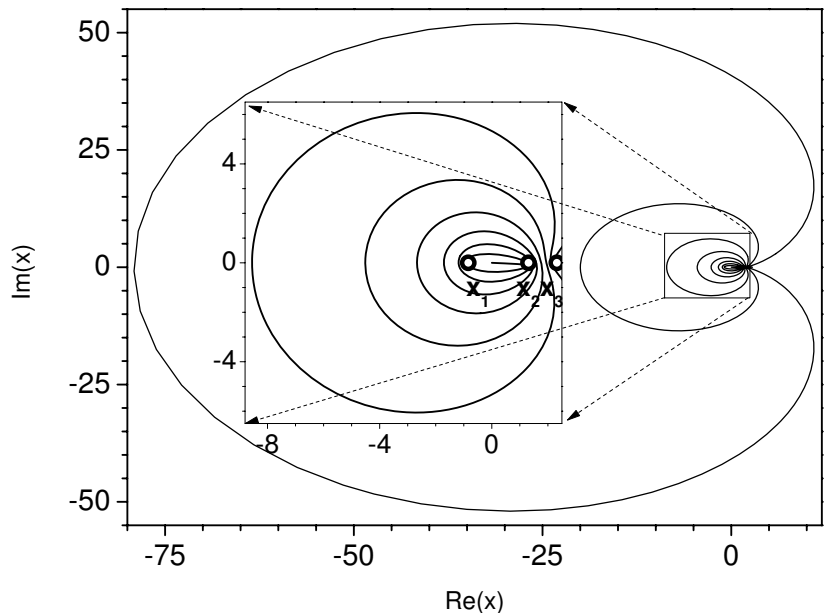}
\end{center}
\caption{The trajectory in the complex x-plane of a classical particle
with complex energy (\ref{b}) at $g=2/\sqrt{125}$. Open circles in the
insertion show the turning points $x_1$, $x_2$ and $x_3$. }
\end{figure}

Recently, in \cite{Bend} it has been argued that "quantum" effects
can be explained in terms of classical trajectories. In
particular, the tunneling effect for a potential
\begin{equation}
V(x)=x^2/2-g x^3 \label{a}
\end{equation}
has been discussed. In this case, the lower state is a quasi-stationary
state whose lifetime $\tau$, defined by the population of the bound
state $P(t)=e^{-t/\tau}$, can be approximated for small $g$ using
WKB theory by the expression \cite{Bend2}
\begin{equation}
\tau=\frac{1}{2}g\sqrt{\pi}\exp(2/15g^2) \label{b}
\end{equation}
In Fig.8 of the paper \cite{Bend}, a complex classical trajectory
with complex energy (\ref{b}) at $g=2/\sqrt{125}$  is represented
(see also insertion in Fig.1 of this paper \footnote{Fig.1 and
Tab.1 have been prepared by Alexander Gusev}) with comment
\\[0.5cm]
"... initially, as the particle crosses the real axis to the right
of the middle turning point, its trajectory is concave leftward,
but as time passes, the trajectory becomes concave rightward. It
is clear that by the fifth orbit the right turning point has
gained control, and we can declare that the classical particle has
now 'tunneled' out and escaped from the parabolic confining
potential. The time at which this classical changeover occurs is
approximately at $t=40$. This is in good agreement with the
lifetime of the quantum state in (8), whose numerical value is
about 20."
\\[0.5cm]
\noindent However, if the time increases further, one can see from
Fig.1 that the trajectory does not escape toward the right side but
continues to rotate mostly on the opposite side. On the other hand, we
can compare the quantum lifetime $\tau$ (\ref{b}), and the classical time
to reach the right turning point $t_c$ -
\begin{equation}
{\rm Re}\ x(t_c)={\rm Re}\ x_3
 \label{c}
\end{equation}
- proposed in \cite{Bend} as a classical analogue of $\tau$. It is
seen from Table 1 that there is nothing in common between these
two quantities in spite of  the "good agreement" declared in
\cite{Bend}. Thus, the effect discussed  is an artificial result
which has no relation to quantum theory.
\begin{table}[tbp]
\caption{The classical time $t_c$ (\ref{c}) and the quantum lifetime
$\tau$ (\ref{b}) versus coupling constant $g$. }
\begin{center}
\begin{tabular}{|r|r|r|r|r|} \hline
 $g$ & 0.12522& 0.14311& 0.16099 & 0.17888 \\ \hline
 $t_c$& 15009 &1385 & 220 & 49 \\
 $\tau$ & 547.3& 85.2 & 24.4& 10.2\\ \hline
\end{tabular}
\end{center}
\end{table}

\section{Concluding remarks}\label{sec:5}

Any theory based on mathematics has internal conflicts. Probably,
only classical mechanics has no evident contradictions. However,
classical electrodynamics is not self-consistent in its background
- an accurate dynamic equation describing the common evolution of
charged particles and the electro-magnetic field does not exist since
the concept 'material point' leads to the divergence of the energy
of its electro-magnetic field, and, on the other hand, the concept
'solid state' is in conflict with relativistic principles. Here we
need a more sophisticated concept than 'material point' and 'solid
state' which is maybe beyond the standard mathematical language.
In quantum mechanics, we cannot clearly formulate a physical
stationary state in the continuum (see Sec.3).

A scientific approach is quite restricted since our vocabulary is
not certain and complete in principle. For instance, it admit such
paradoxes as "Can God create a stone so heavy that he cannot lift
it?".  Another example which demonstrates the incompleteness of
vocabulary is the fact that physics cannot be represented in
Eskimo language where they have dozen of words for different types
of snow (snow which fell down yesterday, snow on which a
dog-sledge passed, {\it etc.}); however, there is no word 'snow'
which is too abstract for them. But what is the level of our
mentality (language $\rightleftharpoons$ mentality)? However it is
the sole tool for communication. In mathematics, the trace of this
incompleteness is G\"odel's theorem \cite{God} which states that
there are true propositions about the naturals that cannot be
proved from the axioms.
\section*{Acknowledgment}
I am grateful to Vladimir Belyaev, John Briggs, Tasko Grozdanov
and participants of the Demkov, Ponomarev and BLTP seminars for
fruitful discussions.

\end{document}